\documentstyle[prd,aps]{revtex} 

\begin{document}    
\input psfig

\title {The collision of two slowly rotating, initially non boosted, black
holes in the close limit.} 
\author{Reinaldo J. Gleiser, and Alfredo E. Dominguez } 
 \address{Facultad de Matem\'atica, Astronom\'\i a y F\'\i sica,  
Universidad Nacional de C\'ordoba \\ 
Ciudad Universitaria,  5000 C\'ordoba, Argentina}
\maketitle

\begin{abstract}

We study the collision of two slowly rotating, initially non boosted, black
holes in the close limit. A ``punctures''  modification of the Bowen - York
method is used to construct conformally flat initial data appropriate to the
problem. We keep only the lowest nontrivial orders capable of giving rise to
radiation of both gravitational energy and angular momentum. We show that even
with these simplifications an extension to higher orders of the linear
Regge-Wheeler-Zerilli black hole perturbation theory, is required to deal with
the evolution equations of the leading contributing multipoles. This extension
is derived, together with appropriate extensions of the Regge-Wheeler and
Zerilli equations. The data is numerically evolved using these equations, to
obtain the asymptotic gravitational wave forms and amplitudes. Expressions for
the radiated gravitational energy and angular momentum are derived and used
together with the results of the numerical evolution to provide quantitative
expressions for the relative contribution of different terms, and their
significance is analyzed.

\end{abstract}
 
\vspace{1cm}

 
\vspace{1cm}

\section{Introduction}

One of the most promising sources of gravitational radiation currently under
investigation is the collision and coalescence of binary pairs of
black holes of comparable mass \cite{flana}. The dynamics of such systems in
the regime corresponding to large separation, where the interaction is weak,
appears to be adequately described in terms of post-Newtonian approximations
\cite{blanchet}. At sufficiently close separation, where the interaction
becomes strong, the system is expected to go through a ``plunge'' phase,
eventually leading to the formation of a common horizon, after which the
system behaves as a perturbed black hole that settles to a final stationary
state. At this latest stage, ``close approximation'' perturbative method have
proved capable of providing reliable information on waveforms and related
quantities \cite{PriPu}. However, up to the present, no equally successful
methods have been found that can handle the intermediate strong field regime,
although some speculations have been presented, based mainly on
extrapolations between the early and very late regimes. The fact that these
methods may not reveal effects that may be unique to the strong field stage of
binary interactions has been recently pointed out by Price and Whelan
\cite{PriWhe}, in relation with the effect of tidal interactions in binary
black hole inspiral, in the case where the black holes have moderately to
large angular momentum. 
 
One of the particular close limit regimes considered in \cite{PriWhe}, that in
which the black holes have equal, parallel spins, pointing in the same
direction and perpendicular to the line joining the centers of the black holes,
had not been analyzed in detail previously in the literature \cite{khanna},
and therefore only rough estimates are given there. In this paper we
provide that analysis by considering the collision of two slowly rotating,
initially non boosted, black holes in the close limit. The black holes have
equal masses and equal parallel spins  aligned perpendicular to the line
joining initially the centers of the black holes. The expression ``initially
non boosted'' refers to the manner in which the initial data is constructed.
In our case we consider initial data obtained by the ``puncture''\cite{BrBr}
modification of the Bowen-York ansatz \cite{BoYo}. Mostly for simplicity, we
include angular, but not linear, momentum in the conformal extrinsic
curvature. In the close limit considered here, this represents a perturbed,
non axially symmetric, state of a (slowly) rotating black hole, and therefore
it is natural to apply the Regge-Wheeler-Zerilli \cite{ReWh},\cite{Zeri}
black hole perturbation theory to analyze its evolution. However, as we
show in this paper, even with the above simplifications the evolution equations
for a part of the leading contributions are linear, but not homogenous, and
require formulating the problem in second order perturbation theory
\cite{physrep}, for a problem with more than one perturbation parameter. The
corresponding evolution equations contain ``source'' terms, and their
influence on the asymptotic behavior of the solutions, which contains the
information on the gravitational waves, needs careful evaluation. Since these
expressions had not been given in the literature before, this paper contains
the details of construction of the initial data and the evolution equations,
together with the corresponding expressions for the gravitational wave
amplitudes, and radiated energy, in terms of (gauge
invariant) Regge-Wheeler-Zerilli functions. 

In the close limit the system
analyzed here bears a marked resemblance with the ``pseudoinspiral
problem'' considered in \cite{psinsp}, \cite{njp}, where the binary black hole
system has orbital, but no spin angular momentum. This resemblance stems from
the fact that in the close limit, once a single horizon has been formed, both
systems behave as distorted black hole settling to a final Kerr form, with
the distortion dominated by quadrupolar terms. Since the systems lack
axial symmetry, it is expected in both cases that the gravitational waves will
carry also  radiation of angular momentum. This expectation was confirmed by
the computations carried out in \cite{psinsp}, \cite{njp} for the
``pseudoinspiral case''. Since only linear perturbation theory is required
in that case, the computations were done both in the Regge-Wheeler-Zerilli
and the Teukolsky formalism \cite{teuko}. However, it was rather surprisingly
found that although the radiated energy computed in both formalism showed
a reasonable agreement, this did not happen for the predicted radiation of
angular momentum. In fact, except for very small values of the angular
momentum, the Teukolsky formalism, as applied in \cite{psinsp}, \cite{njp},
leads to the rather disturbing result that the angular momentum increases as a
result of the radiation (the angular momentum radiated has opposite sign to
that of the system). This is not the case for the prediction given by the
Regge-Wheeler formalism. Although the detailed reasons for this discrepancy are not
clearly understood, it is the opinion of the present authors that it may be
due mainly to the rather large departure of the initial data from that of a
Kerr black hole. Since the system considered in this paper has initial
data that is closer to that of a Kerr black hole, in the sense that the
perturbations are of second order in the angular momentum parameter, rather
than first order as in the case of \cite{psinsp}, \cite{njp}, one of the main 
purposes of our analysis was to compute the angular momentum radiated in the
process, as the disturbed black hole settles to its final Kerr form, and its
eventual comparison with the corresponding Teukolsky results. As is shown in
this paper, within the Regge-Wheeler-Zerilli formalism, this, as well as the
computation of the leading wave forms and radiated energy, requires an
extension of the theory to second order, (in fact, to even higher order, if we
derive the expression for the radiated angular momentum as indicated in
Section VII), which is explicitly given here. It should be clear that at
this point that the alternative treatment be based on the Teukolsky
formalism, and, possibly, its extension to second order \cite{campa},
is a task requiring as much extension as the one reported here. For this
reason, that development, and its comparison with the results obtained with
the Regge-Wheeler formalism presented here, will be considered in a separate
paper, currently under development.

The paper is organized as follows. In the next Section we consider the
construction of the initial data. This is naturally given in a ``Misner'' type
gauge. However, as we indicate, this gauge is not appropriate for a
perturbative computation, and we derive the gauge transformation to, and the 
form of the initial data in a ``Boyer-Lindquist-Kerr'' type of gauge for the
$\ell=1$, odd parity perturbations, where these have a very simple coordinate
dependence. We then consider the evolution of the second order perturbations
in the Regge-Wheeler gauge, and obtain the corresponding Zerilli, and
Regge-Wheeler functions and equations. We describe their asymptotic behavior,
their relation to the corresponding gravitational wave amplitudes, and derive
the expressions for the radiated power and energy. Finally, we obtain an
expression for the radiated angular momentum directly using perturbation
theory, in terms of the Zerilli functions for the perturbations. Comments
and conclusions are given in the last Section.

\section{Initial data}

In applications of the Regge - Wheeler - Zerilli black hole perturbation theory
one considers a family of metrics, depending on one or a few parameters, such
that the Schwarzschild metric is recovered when the parameters are set equal to
zero. The problem may be analyzed as an initial value problem, so that the
specification of the members of the family amounts to identifying the
corresponding initial data on a certain three dimensional hypersurface.
Actually, since the computations are carried out in a coordinate patch that
covers only the region outside the black hole horizon, one requires expansions
of the initial that may be convergent only in that region. Examples are
furnished by the families of initial data constructed via the ``conformal
approach'' \cite{BoYo} to the initial value problem in General Relativity. In
this approach one assumes that the three-metric $g_{ab}$ is conformally flat,
i.e. $g_{ab}= \Phi^4 \eta_{ab}$, where $\Phi$ is some non vanishing function,
and $\eta_{ab}$ a flat three-metric, and defines a conformal extrinsic
curvature $\widehat{K}_{ab}= \Phi^2 K_{ab}$. Then, assuming maximal slicing
$K^{ab} g_{ab}=0$, the initial value constraint (for a vacuum metric) are
reduced to, 
\begin{eqnarray} 
\label{momentum} 
\nabla_a \widehat{K}^{ab} & = & 0 \\ 
\label{hamiltonian} 
\nabla^2 \Phi & = &- {1 \over 8} {\widehat{K}^{ab}\widehat{K}_{ab} \over \Phi^7} 
\end{eqnarray}
where all derivatives are with respect to the flat metric. In general, the
conditions,
\begin{equation}
\label{asymp01} 
  \Phi > 0 
\;\; , \;\; \lim_{\rho \rightarrow \infty} \Phi = 1.   
\end{equation}
where $\rho$ is a radial coordinate on the fiducial flat space where $\eta$
is defined, are imposed on $\Phi$. Therefore, on the initial slice,
asymptotically, for large $\rho$ we have $\widehat{K}^{ab} \rightarrow
{K}^{ab}$, and $g_{ab} \rightarrow \eta_{ab}$. This fact is important because
one can find exact solutions to the momentum constraint (\ref{momentum}), and,
using (\ref{asymp01}), obtain a (partial) physical interpretation of the 
associated spacetime in terms of ADM observables, without solving
(\ref{hamiltonian}) explicitly. For example, in \cite{singlespin}, the solution
\begin{equation}
\label{K_one} 
\widehat{K}_{ab} = {3 \over \rho^3} \left[  \epsilon_{acd} S^c n^d n_b + n_a 
\epsilon_{bcd} S^c n^d  \right],
\end{equation}
where $\rho = \sqrt{(x^1)^2+(x^2)^2+(x^3)^2}$, $n^a=x^a/\rho$, with $x^a$
cartesian coordinates on the flat space background, and $\bf{S}$ is a constant
vector, is considered. Then, taking spherical polar coordinates for the flat
background, so that
\begin{equation}
\eta_{ab} dx^a dx^b = (d\rho)^2+\rho^2(d\theta)^2 + \rho^2 \sin(\theta)^2 (d
   \phi)^2 
\end{equation}
one straightforwardly identifies $S^a$ with the angular momentum of the
system, irrespective of the detailed knowledge of $\Phi$. In the application
given in \cite{singlespin}, the conditions (\ref{asymp01}) are supplemented
with appropriate boundary conditions for $\Phi$, so that the resulting initial
data is identified with that of a single spinning black hole (in a
non stationary initial state), centered at the origin of the flat coordinate
system.

In this paper we consider the initial data that results from taking
$\widehat{{K}}_{ab}$ as the sum of two terms of the form (\ref{K_one}),  
symmetrically centered around the coordinate origin, with the same $S^a$. In
the cartesian coordinate system associated with the flat background, we have,
\begin{eqnarray}
\label{K_two} 
\widehat{K}_{ab} & = & {3 \over \rho^3_1} \left[  \epsilon_{acd} S^c n_1^d
n_{1\;b}  + n_{1\;a} \epsilon_{bcd} S^c n_1^d  \right] \nonumber \\
& & +  {3 \over \rho^3_2} \left[  \epsilon_{acd} S^c n_2^d n_{2\;b} + 
n_{2\;a} \epsilon_{bcd} S^c n_2^d  \right]
\end{eqnarray}
where $\rho_1 =  \sqrt{(x^1-L/2)^2+(x^2)^2+(x^3)^2}$, $\rho_2 = 
\sqrt{(x^1+L/2)^2+(x^2)^2+(x^3)^2}$, $n^a_1=x^a/\rho_1$, and
$n^a_2=x^a/\rho_2$. We also take $S^a=(0,0,S)$. This may be thought of as the
(conformal) extrinsic curvature corresponding to two black holes with equal
angular momentum $S$, pointing along the $z$-axis, placed at the points $(\pm
L/2,0,0)$. Actually, this interpretation requires that $\Phi$ has a 
singularity structure in accordance with that of (\ref{K_two}). This may be
achieved in different ways. Here we take the ``punctures'' ansatz \cite{BrBr},
namely, we assume that $\Phi$ is of the form,
\begin{equation}
\label{punkt1}
\Phi = \Phi_{BL} +\Phi_{Reg}
\end{equation}
where,
\begin{equation}
\label{punkt2}
\Phi_{BL}= 1 + {M_0 \over 2 \rho_1} + {M_0 \over 2 \rho_2}
\end{equation}
that is, $\Phi_{BL}$ is taken as the Brill-Lindquist conformal factor
\cite{brilin}, and we impose that $\Phi_{Reg}$ must be regular in the whole
conformal plane, and vanish for large $\rho_i$. In accordance with this
prescription, $\Phi_{Reg}$ satisfies the equation,
\begin{equation}
\label{punkt3}
\nabla^2 \Phi_{Reg} = - {1 \over 8} {\widehat{K}^{ab}\widehat{K}_{ab} \over 
( \Phi_{BL} +\Phi_{Reg})^7}
\end{equation}
with $\widehat{K}^{ab}$ given by (\ref{K_two}), which, for $S \neq 0$, 
may only be solved  numerically. We notice, however, that for $S=0$ we should
have $\Phi_{Reg}=0$, and, from the form (\ref{punkt3}), we expect that
$\Phi_{Reg}$ admits an expansion in powers of $S^2$, near $S=0$.  Moreover,
when $L \rightarrow 0$, $\Phi_{BL}$ approaches the conformal factor for a
Schwarzschild black hole. Therefore, the ansatz (\ref{K_two}), together with
(\ref{punkt1}), and (\ref{punkt2}), lead to a two-parameter family of initial
data, with the Schwarzschild metric as the limit when the parameters vanish.
In the slow, close approximation approach of this paper we consider the
situation when the two holes are initially so close to each other that a
single horizon is formed, and the system, from the point of view external to
the horizon, may be considered as a perturbed single black hole. In this
situation, the external field admits a multipolar expansion, and the
multipolar order turns out to be correlated with different powers of the
parameters, so that, by keeping the lowest orders in the parameters, we end up
with a few multipolar terms, and this is expected to provide a good
approximation to the exact initial data, and its evolution. 

There is always a certain degree of arbitrariness in the
choice of perturbation parameters. Here we notice that, because of the
conditions imposed on $\Phi$, the initial data corresponds to a system with
ADM angular momentum $J= 2 S$. We shall therefore consider $J$, and $L$ as
expansion parameters. In the limit $J=0$, $L=0$, the conformal factor $\Phi$
corresponds to that of a Schwarzschild black hole of mass $M=2 M_0$. In what
follows we write all expressions in terms of $M$, rather than $M_0$. 

\subsection{Perturbative expansion of the initial data}

We recall that in the Regge-Wheeler black hole perturbation theory the
analysis is restricted to the region outside the horizon. To construct our
perturbative initial data we start by considering (\ref{punkt3}), but
restricted to the region $\rho > L$. If we expand
$\widehat{K}_{ab}$ in powers of $L/\rho$, and keep only lowest orders, we find,

\begin{eqnarray}
\label{Khat1}
\widehat{K}_{\rho \rho} & = &  -3 {J L^2 \sin(2 \phi) \sin^2(\theta) \over 
\rho^5} \nonumber \\
\widehat{K}_{\rho \theta} & = & -{9 \over 8} { J L^2 \sin(2 \phi) \sin
(\theta)  \cos(\theta) \over \rho^4} \nonumber \\
\widehat{K}_{\rho \phi} & = & 3 {J \sin^2(\theta) \over \rho^2} 
- {3 \over 8}{J L^2 (7 + \cos^2(\phi)(25 \cos^2(\theta)-19)) \sin^2(\theta)
\over \rho^4} \nonumber \\ 
\widehat{K}_{\theta \theta} & = & {3 \over 4} {J L^2 \sin(2 \phi)
\cos^2(\theta)/\rho^3} \nonumber \\ 
\widehat{K}_{\theta \phi} & = & - {3 \over 4} {J L^2
\sin(\theta) \cos(\theta) (1 +3 \cos^2(\phi)-5 \cos^2(\phi) \cos^2(\theta)
\over \rho^3} \nonumber \\ 
\widehat{K}_{\phi \phi} & = & {3 \over 4} {J L^2
\sin(2\phi)(4-9 \cos^2(\theta)+5 \cos^4(\theta) \over \rho^3}    
\end{eqnarray}

This implies, keeping again only the lowest contributing order,
\begin{equation}
\label{Khatsq}
\widehat{K}_{ab}\widehat{K}^{ab} = 18 J^2 {\sin^2(\theta) \over \rho^6}  
\end{equation}

Similarly, we expand $\Phi_{BL}$,
\begin{equation}
\label{exfiBL}
\Phi_{BL} = 1 + {M \over  2 \rho}- { M(1-3 \cos^2(\phi) 
\sin^2(\theta))L^2 \over 16 \rho^3} 
\end{equation}

A simple power counting shows that the lowest order in $\Phi_{Reg}$ is $J^2$,
if we consider $J^2$ and $L^2$ of the same order. To this order (\ref{punkt3})
reduces to,
\begin{equation}
\label{punkt5}
\nabla^2 \Phi_{Reg} = - 288 J^2 {\sin^2(\theta) \rho\over (2 \rho+M)^7}
\end{equation}
 
The solution of this equation is,
\begin{equation}
\label{punkt6}
 \Phi_{Reg} =  \Phi_{(0,0)}(\rho) Y_0^0(\theta,\phi)+\Phi_{(2,0)}(\rho) 
Y_2^0(\theta,\phi)
\end{equation}
where $Y_{\ell}^m(\theta,\phi)$ is a standard spherical harmonic, and,
\begin{eqnarray}
\label{punkt7}
\Phi_{(0,0)}(\rho)  & = &  { 4 \sqrt{\pi} J^2 (M^4+10 \rho M^3
+40 \rho^2 M^2+40\rho^3 M+ 16\rho^4) \over 5 M^3(2 \rho+M)^5} + C_1
+{C_2\over\rho}  \nonumber \\  
\Phi_{(2,0)}(\rho)  & = &  { \sqrt{ 5 \pi}
J^2 (M^4+10 \rho M^3+40 \rho^2 M^2+80\rho^3 M+80\rho^4) 
\over 25\rho^3(2 \rho+M)^5} + C_3 \rho^2 +{C_4\over\rho^3}     
 \end{eqnarray}

We must set $C_1=C_3=0$, to have the correct asymptotic behavior. The
constants $C_2$, and $C_4$ are fixed so that $\Phi_{(0,0)}$, and 
$\Phi_{(2,0)}$ are regular for $\rho=0$. This, for our perturbative
expansion,  is a consequence of the regularity conditions imposed on
$\Phi_{Reg}$ within the ``punctures'' ansatz \cite{BrBr}. See also
the discussion given in \cite{boosted}, for the analogous
case of ``boosted'' black hole collisions. The final result is, 

\begin{eqnarray}
\label{punkt8}
\Phi_{(0,0)}(\rho)  & = &  { 4 \sqrt{\pi} J^2 (M^4+10 \rho M^3
+40 \rho^2 M^2+40\rho^3 M+ 16\rho^4) \over 5 M^3(2 \rho+M)^5}
  \nonumber \\
 \Phi_{(2,0)}(\rho)  & = & - {32 \sqrt{ 5 \pi} J^2 \rho^2 
\over 25 M (2 \rho+M)^5}      
\end{eqnarray}

This completes our perturbative computation of the conformal factor and
extrinsic curvature. To construct the initial data for the evolution
equations, we recall that if we fix the shift functions $N_a=0$, we have
\begin{eqnarray}
\label{metric1}
ds^2 & = & g_{ab} dx^a dx^b - N^2 dt^2 \nonumber \\
{\partial g_{ab} \over \partial t} & = & -2 N K_{ab}
\end{eqnarray}
where for $t=0$ we identify $g_{ab}$, and $K_{ab}$ with those obtained above.
We next change from the conformal radial coordinate $\rho$ to a Schwarzschild
radial coordinate $r$, such that $\rho=(\sqrt{r}+\sqrt{r-2M})^2/4$, and choose
$N=\sqrt{1-2M/r}$, so that we recover the Schwarzschild
metric, with mass $M$, for $J=L=0$. The final results, in the standard
Regge-Wheeler multipolar decomposition and notation \cite{ReWh} (see. e.g.
\cite{physrep}, for more details) are listed below. Notice, however,
that we indicate the odd parity perturbation
components with $k0_{\ell,m}$, $k1_{\ell,m}$ and $k2_{\ell,m}$,
instead of $h0_{\ell,m}$, $h1_{\ell,m}$ and $h2_{\ell,m}$, as in
\cite{ReWh} to avoid the explicit use of ``odd'' and ``even'' labels.

\subsection{Metric components}

The metric components obtained using the procedure just described may be
written as follows,

\subsubsection{ $\ell =0$}

\begin{equation}
{{H2}_{0, \,0}} = {{K}_{0, \,0}} ={\displaystyle \frac {4}{5}} \,
{\displaystyle \frac {\sqrt{\pi }\,J^{2}\,(r\,M + M^{2} + 2\,r^{2
})}{r^{3}\,M^{3}}}  
\end{equation}

\subsubsection{ $\ell =2$} 

\begin{equation}
{{H2}_{2, \,0}} = {{K}_{2, \,0}} = - {\displaystyle \frac {4}{25}} \,
{\displaystyle \frac {\sqrt{5}\,\sqrt{\pi }\,J^{2}}{M\,r^{3}}} 
 - {\displaystyle \frac {16}{5}} \,{\displaystyle \frac {M\,
\sqrt{5}\,\sqrt{\pi }\,L^{2}}{\sqrt{r}\,(\sqrt{r} + \sqrt{r - 2\,
M})^{5}}} 
\end{equation}

\begin{equation}
{{H2}_{2, \,-2}} = {{H2}_{2, \,2}} = {\displaystyle \frac {8}{5}} \,
{\displaystyle \frac {\sqrt{\pi }\,L^{2}\,M\,\sqrt{30}}{\sqrt{r}
\,(\sqrt{r} + \sqrt{r - 2\,M})^{5}}} 
\end{equation}
 
\begin{equation} 
{{K}_{2, \,-2}} = {{K}_{2, \,2}} = {\displaystyle \frac {8}{5}} \,
{\displaystyle \frac {\sqrt{\pi }\,L^{2}\,M\,\sqrt{30}}{\sqrt{r}
\,(\sqrt{r} + \sqrt{r - 2\,M})^{5}}} 
\end{equation}
 
\begin{eqnarray}
{{\partial_t H2}_{2, \,-2}} & = & -{{\partial_t H2}_{2, \,2}} =
{\displaystyle \frac {64}{5}} \, {\displaystyle \frac {i\,J\,\sqrt{\pi
}\,L^{2}\,\sqrt{r - 2\,M}\, \sqrt{30}}{r^{(7/2)}\,(\sqrt{r} + \sqrt{r -
2\,M})^{4}}}  \\
{{\partial_t h1}_{2, \,-2}} & = &  - {\displaystyle \frac {8}{15}} 
\,{\displaystyle \frac {i\,J\,(8\,r + 8\,\sqrt{r}\,\sqrt{r - 2\,M
} - 3\,M)\,\sqrt{\pi }\,L^{2}\,\sqrt{30}}{r^{(5/2)}\,(\sqrt{r} + 
\sqrt{r - 2\,M})^{5}}} 
\\
{{\partial_t h1}_{2, \,2}} & = & - {{\partial_t h1}_{2, \,-2}}
\\
{{\partial_t G}_{2, \,-2}} & = & -{{\partial_t G}_{2, \,2}} =
{\displaystyle \frac {16}{15}} \, {\displaystyle \frac {i\,J\,\sqrt{\pi
}\,L^{2}\,\sqrt{r - 2\,M}\, \sqrt{30}}{r^{(7/2)}\,(\sqrt{r} + \sqrt{r -
2\,M})^{4}}}  
\\
{{\partial_t K}_{2, \,-2}} & = & -{{\partial_t K}_{2, \,2}} = -
{\displaystyle \frac {16}{5}} \, {\displaystyle \frac {i\,J\,\sqrt{\pi
}\,L^{2}\,\sqrt{r - 2\,M}\, \sqrt{30}}{r^{(7/2)}\,(\sqrt{r} + \sqrt{r -
2\,M})^{4}}}  
\end{eqnarray}

\subsubsection{ $\ell =1$, odd}

\begin{equation}
{{\partial_t k1}_{1, \,0}} = 4\,{\displaystyle \frac {\sqrt{3}\,
J\,\sqrt{\pi }}{r^{2}}}  - {\displaystyle \frac {16}{5}} \,
{\displaystyle \frac {\sqrt{3}\,J\,\sqrt{\pi }\,L^{2}\,M}{(\sqrt{
r} + \sqrt{r - 2\,M})^{5}\,r^{(5/2)}}} 
\end{equation}

\subsubsection{ $\ell =3$, odd}

\begin{eqnarray}
{{\partial_t k1}_{3, \,-2}} & = & {\displaystyle \frac {4}{105}} \,
{\displaystyle \frac {J\,L^{2}\,\sqrt{\pi }\,\sqrt{210}(25\,r + 25\,
\sqrt{r}\,\sqrt{r - 2\,M}
 - 6\,M)}{r^{(5/2)}\,(\sqrt{r} + 
\sqrt{r - 2\,M})^{5}}} 
\\
{{\partial_t k1}_{3, \,0}} & = &  - {\displaystyle \frac {8}{35}} 
\,{\displaystyle \frac {\sqrt{\pi }\,\sqrt{7}\,L^{2}\,J\,(25\,r
 + 25\,\sqrt{r}\,\sqrt{r - 2\,M} - 6\,M)}{(\sqrt{r} + \sqrt{r - 2
\,M})^{5}\,r^{(5/2)}}} 
\\
{{\partial_t k1}_{3, \,2}} & = & {\displaystyle \frac {4}{105}} \,
{\displaystyle \frac {J\,L^{2}\sqrt{\pi }\,\sqrt{210}\,(25\,r + 25\,
\sqrt{r}\,\sqrt{r - 2\,M}
 - 6\,M) }{r^{(5/2)}\,(\sqrt{r} + 
\sqrt{r - 2\,M})^{5}}} 
\\
{{\partial_t k2}_{3, \,-2}} & = & {{\partial_t k2}_{3, \,2}} = {\displaystyle
\frac {8}{21}} \, {\displaystyle \frac {J\,\sqrt{\pi }\,L^{2}\,\sqrt{r -
2\,M}\, \sqrt{210}}{r^{(3/2)}\,(\sqrt{r} + \sqrt{r - 2\,M})^{4}}} 
\\
{{\partial_t k2}_{3, \,0}} & = &  - {\displaystyle \frac {16}{7}} \,
{\displaystyle \frac {\sqrt{\pi }\,\sqrt{r - 2\,M}\,L^{2}\,J\,
\sqrt{7}}{(\sqrt{r} + \sqrt{r - 2\,M})^{4}\,r^{(3/2)}}} 
\end{eqnarray} 
  
\subsection{Extrinsic curvature components}

The  extrinsic curvature  components may be written in the
form,

\subsubsection{$\ell=1$,odd}

\begin{equation}
{{Kk1}_{1, \,0}} =  - 4\,{\displaystyle \frac {\sqrt{3}\,
J\,\sqrt{\pi }}{r^{(3/2)}\,\sqrt{r - 2\,M}}}  + {\displaystyle 
\frac {16}{5}} \,{\displaystyle \frac {\sqrt{3}\,J\,\sqrt{\pi }\,
L^{2}\,M}{(\sqrt{r} + \sqrt{r - 2\,M})^{5}\,r^{2}\,\sqrt{r - 2\,M
}}} 
\end{equation}

\subsubsection{$\ell=2$, even}

\begin{eqnarray}
{{KH2}_{2, \,-2}} & = & -{{KH2}_{2, \,2}} =  - {\displaystyle \frac
{32}{5}} \, {\displaystyle \frac {i\,L^{2}\,J\,\sqrt{30}\,\sqrt{\pi }}{(
\sqrt{r} + \sqrt{r - 2\,M})^{4}\,r^{3} }} 
\\
{{Kh1}_{2, \,-2}} & = & -{{Kh1}_{2, \,2}} = {\displaystyle \frac {4}{15}} \,
{\displaystyle \frac {i\,J\,(8\,r + 8\,\sqrt{r}\,\sqrt{r - 2\,M}
 - 3\,M)\,\sqrt{\pi }\,L^{2}\,\sqrt{30}}{r^{2}\,\sqrt{r - 2\,M}\,
(\sqrt{r} + \sqrt{r - 2\,M})^{5}}} 
\\
{{KG}_{2, \,-2}} & = & -{{KG}_{2, \,2}} = - {\displaystyle \frac
{8}{15}} \, {\displaystyle \frac {i\,J\,\sqrt{\pi }\,L^{2}\,\sqrt{30}}{r^{3}
\,(\sqrt{r} + \sqrt{r - 2\,M})^{4}}} 
\\
{{KK}_{2, \,-2}} & = & - {{KK}_{2, \,2}} = {\displaystyle \frac {8}{5}} \,
{\displaystyle \frac {i\,J\,\sqrt{\pi }\,L^{2}\,\sqrt{30}}{r^{3}
\,(\sqrt{r} + \sqrt{r - 2\,M})^{4}}} 
\end{eqnarray}

\subsubsection{$\ell=3$, odd}

\begin{eqnarray}
{{Kk1}_{3, \,-2}} & = & {{Kk1}_{3, \,0}} =  - {\displaystyle \frac {2}{105}} \,
{\displaystyle \frac {J\,(25\,r + 25\,\sqrt{r}\,\sqrt{r - 2\,M}
 - 6\,M)\,\sqrt{\pi }\,L^{2}\,\sqrt{210}}{r^{2}\,\sqrt{r - 2\,M}
\,(\sqrt{r} + \sqrt{r - 2\,M})^{5}}} 
\\
{{Kk1}_{3, \,0}} & = & {\displaystyle \frac {4}{35}} \,
{\displaystyle \frac {\sqrt{\pi }\,\sqrt{7}\,L^{2}\,J\,(25\,r + 
25\,\sqrt{r}\,\sqrt{r - 2\,M} - 6\,M)}{(\sqrt{r} + \sqrt{r - 2\,M
})^{5}\,r^{2}\,\sqrt{r - 2\,M}}} 
\\
{{Kk2}_{3, \,-2}} & = &  {{Kk2}_{3, \,2}} =  - {\displaystyle \frac {4}{21}} \,
{\displaystyle \frac {J\,\sqrt{\pi }\,L^{2}\,\sqrt{210}}{r\,(
\sqrt{r} + \sqrt{r - 2\,M})^{4}}} 
\\
{{Kk2}_{3, \,0}} & = & {\displaystyle \frac {8}{7}} \,
{\displaystyle \frac {\sqrt{\pi }\,L^{2}\,J\,\sqrt{7}}{(\sqrt{r}
 + \sqrt{r - 2\,M})^{4}\,r}} 
\end{eqnarray}
where we are using our modified Regge-Wheeler notation, and a $K$ at
the beginning of each function indicates that we have a multipolar
component of the extrinsic curvature, instead of one of the metric.

\section{The choice of gauge for the $\ell=1$, odd parity, perturbation}

If we consider the initial data for the $\ell=1$, odd, perturbation, we find
that if we set $k0_{1,0}=0$, consistent with setting the shifts to zero, the
evolution equations have the solution,
\begin{equation}
{ k1}_{1, \,0} = 4\,{\displaystyle \frac {\sqrt{3}\,
J\,\sqrt{\pi } t }{r^{2}}}  
\end{equation}

This in turn leads to solutions for the other amplitudes that grow in time.
For our calculations it will be more convenient to choose a gauge where,
\begin{eqnarray}
k0_{1,0} & = & { 4 \sqrt{\pi}J \over \sqrt{3} r} \\
k1_{1,0} & = & 0
\end{eqnarray} 

The gauge transformation required to produce this change is first order in $J$,
for the $\ell=1$, odd parity terms, and introduces changes in the remaining
terms of order $J^2$, and $J L^2$.  A simple computation,
using second order perturbation theory, as given e.g. in \cite{physrep}, shows
that for the $\ell=2$, even parity, order $J^2$ terms, the initial data
resulting from this change is the same as for the initial gauge except that
now we have,  
\begin{equation} 
{H0}_{2, \,0} =  - {16  \sqrt{5 \pi} J^2 \over 15 r^3(r-2 M)} 
\end{equation}

Similarly, the transformation induces changes in the $\ell=2$, even
parity, order $J L^2$ terms. In this case, these correspond to first
order perturbation, because the terms are linear in $J$. A
straightforward computation shows that the expressions derived
previously are replaced by,  
\begin{eqnarray} {{\partial_t H2}_{2,
\,2}} & = & -{{\partial_t H2}_{2, \,-2}} =- {\displaystyle \frac {16}{5
}} \,{\displaystyle \frac {i\,\sqrt{\pi }\,\sqrt{30}\,(\sqrt{r
} + 3\,\sqrt{r - 2\,M})}{r^{(7/2)}\,(\sqrt{r} + \sqrt{r - 2\,M})
^{4}}} 
\\
{{h0}_{2, \,2}} & = - & {{h0}_{2, \,-2}} ={\displaystyle \frac
{16}{15} } \,{\displaystyle \frac {i\,\sqrt{\pi
}\,M\,\sqrt{30}}{r^{(3/2)} \,(\sqrt{r} + \sqrt{r - 2\,M})^{5}}}    
\\
{{\partial_t h1}_{2, \,2}} & = & - {{\partial_t h1}_{2, \,-2}}
={\displaystyle \frac {8}{15}} 
\,{\displaystyle \frac {i\,\sqrt{\pi }\,\sqrt{30}\,(8\,r + 8\,
\sqrt{r}\,\sqrt{r - 2\,M} - 9\,M)}{(\sqrt{r} + \sqrt{r - 2\,M})^{
5}\,r^{(5/2)}}}  
\\
{{\partial_t G}_{2, \,2}} & = & - {{\partial_t G}_{2, \,-2}}= 
- {\displaystyle \frac {16}{15
}} \,{\displaystyle \frac {i\,\sqrt{\pi }\,\sqrt{30}\,\sqrt{r
 - 2\,M}}{r^{(7/2)}\,(\sqrt{r} + \sqrt{r - 2\,M})^{4}}}  
\\
{{\partial_t K}_{2, \,2}} & = & -{{\partial_t K}_{2, \,-2}} =  
{\displaystyle \frac {16}{5}} 
\,{\displaystyle \frac {i\,\sqrt{\pi }\,\sqrt{30}\,( - 4\,M + 
\sqrt{r}\,\sqrt{r - 2\,M} + r)}{r^{(7/2)}\,(\sqrt{r} + \sqrt{r - 
2\,M})^{5}}} 
\end{eqnarray}
where we now have a non vanishing shift term ($ h0_{2,\pm2} \neq
0$), and the data not explicitly indicated is zero.

Finally, the gauge transformation equations for the $\ell=3$, $m=0, 
\pm 2$, odd
parity perturbations take the form, 
\begin{eqnarray}
\widetilde{k0}_{3,0} & = & {2  \over \sqrt{35} r} \left[K^{L^2}_{2,0}
-2 G^{L^2}_{2,0} \right] \nonumber \\
\widetilde{k1}_{3,0} & = & {k1}_{3,0} -{6 t \over \sqrt{35} r^2}
\left[K^{L^2}_{2,0} -2 G^{L^2}_{2,0} \right] \nonumber \\
\widetilde{k2}_{3,0} & = & k2_{3,0}   \nonumber \\
\widetilde{k0}_{3,\pm 2} & = & {2 \sqrt{7} \over 21 r}
\left[K^{L^2}_{2,\pm 2} -2 G^{L^2}_{2,\pm 2} \right] \nonumber \\
\widetilde{k1}_{3,\pm 2} & = & {k1}_{3,\pm 2} -{2 t \over \sqrt{7}
r^2} \left[K^{L^2}_{2,\pm 2} -2 G^{L^2}_{2,\pm 2} \right] \nonumber
\\ \widetilde{k2}_{3,\pm 2} & = & k2_{3,\pm 2},   
 \end{eqnarray}
where $K^{L^2}_{2,\pm 2}$, $ G^{L^2}_{2,\pm 2}$ are the even parity,
order $L^2$, Regge - Wheeler functions in the Bowen - York gauge. In
general, if necessary, in this paper we shall use a
superindex $L^2$, or $L^2 J$, to make explicit the order of the terms.
The above equations may be used to give explicit expressions for the
initial data in terms of $r$. We shall not display them here.

\section{The evolution equations for the initial data}

The time evolution of the initial data we have just constructed may be 
handled as follows. First we notice that the initial data depends on two
parameters: $J$,and $L$. Actually, only even powers of $L$ appear. This
implies that, in a Regge-Wheeler type of expansion, terms of order $J$, or of
order $L^2$ satisfy linear perturbation equation. We will, therefore, consider
$J$, and $L^2$ as our perturbation parameters, and will restrict to the
leading, non trivial, contributing orders and multipoles. In our case they
may be classified as follows. a) Terms of order $L^2$. These correspond to
the $\ell=2$, even parity, multipole, with $m=0, \pm2$, and satisfy linear
perturbation equations. b) Terms of order $J$ that correspond to the $\ell=1$,
odd, multipole with $m=0$, and are related to the angular momentum. Their time
evolution is known explicitly. c) Terms of order $J^2$, with $\ell=2$, even
parity, $m=0$. These are second order in $J$, and obey linear equations with a
``source'', and may be handled as in \cite{singlespin}, but here we
shall use second order perturbation theory to analyze their evolution. d)
Terms of order $J L^2$. These are actually higher order in perturbation
theory.  Since terms of order $J$ have $\ell=1$, and are of odd parity, and
terms of order $L^2$  have $\ell=2$, and are of even parity, in principle,
they may contribute ``sources'' for terms of order $J L^2$, and $\ell =
1,2,3$, with both even and odd parity. However, an analysis of the form of the
corresponding Einstein equations shows that the perturbations corresponding to
$\ell=1,3$, with even parity, and $\ell =2$ with odd parity, obey linear
homogeneous equations. Since the initial data for these terms vanishes, they
vanish for all times. Of the remaining terms, those with $\ell=1$ odd parity
do not contribute to the radiation, and therefore, shall not be considered in
what follows.  In the next subsections we obtain and display the relevant
equation for the evolution of the terms contributing to the  radiation of
angular momentum and energy, up to the orders considered. This implies, in
some cases, that we shall have to extend previous results on the application
of black hole perturbation theory in the Regge-Wheeler-Zerilli framework. 
These will be indicated where appropriate.

\subsection{Order $L^2$}

In this case we have to consider terms of even parity with $\ell=2$, and
$m=0,\pm2$. The reality of
the metric implies that the amplitudes with $m=-2$ are the complex
conjugate of the corresponding amplitude with $m=+2$, so we need to consider
only one of these.  Since all these perturbations obey first order equations,
in the Regge-Wheeler gauge they may be uniquely written in terms of  the
Zerilli functions $\psi _{2, \,m}(t, \,r)$, $m= 0, \pm 2$,  in the form
\cite{physrep},

\begin{eqnarray}
\label{ele2L2}
{H0}_{2, \,m}(t, \,r) & = &{H2}_{2, \,m}(t, \,r) \nonumber \\
{H2}_{2, \,m}(t, \,r) & = & (r - 2\,M)\,{\partial^2 \psi_{2, \,m}(t,
\,r)\over \partial r^2}+{(2r^2-2rM+3M^2)\over r (2r+3M)}{\partial \psi_{2,
\,m}(t, \,r)\over \partial r} \nonumber \\ 
 & &   - 3\,{\displaystyle \frac { (3\,M^{3} + 6\,r\,M^{2} + 4\,r^{2}\,M +
4\,r^{3})}{(2\,r + 3\,M )^{2}\,r^{2}}} {\psi _{2, \,m}}(t, \,r)  \nonumber \\
 {K}_{2, \,m}(t, \,r) & = & 
 {\displaystyle \frac {(6\,r^{2} + 6\,r\,M + 6\,M^{2}) }{r^{2}\,(2\,r +
3\,M)}}{\psi
 _{2, \,m}}(t, \,r)  + (1 - 2\, {\displaystyle \frac {M}{r}} )\,({\frac {\partial
}{\partial r}} \,{\psi _{2, \,m}}(t, \,r)) \nonumber \\ 
 {H1}_{2, \,m}(t, \,r) & = &{\displaystyle \frac {(2\,r^{2}
 - 6\,r\,M - 3\,M^{2}) }{(r - 2\,M)\,(2\,r + 3\,M)}} \,{\partial \psi _{2, \,m}(t, \,r)
 \over \partial t}  + r\,({\frac {
\partial ^{2}}{\partial t\,\partial r}}\,{\psi _{2, \,m}}(t, \,r)
)
\end{eqnarray}

The Zerilli functions $\psi _{2, \,m}(t, \,r)$ are given in a general gauge by
the Moncrief \cite{moncri} expression,
\begin{eqnarray}
\label{Zerfunl2L}
\psi _{2, \,m}(t, \,r) & = & {r(r-2M) \over 3 (2r+3M)}\left[{H2}_{2, \,m}- r
{\partial {K}_{2, \,m}   \over \partial r} - {r-3M \over r-2M} {K}_{2,
\,m}\right] \nonumber \\
& & + {r^2 \over (2r+3M)} \left[{K}_{2, \,m}+(r-2M) \left( {\partial {G}_{2,
\,m}   \over \partial r} - {2 \over r^2} {h1}_{2, \,m}\right) \right]
\end{eqnarray}
and obey the equation,
\begin{equation}
\label{Zereql2L}
{\partial^2  \psi _{2, \,m} \over \partial t^2} = {\partial^2 \psi _{2, \,m}
\over \partial {r^*}^2}- 6 {(r - 2 M)(3 M^3 + 6 r M^2 + 4r^2 M +
4r^3)\over r^4 (2 r + 3 M)^2}\,\psi _{2, \,m}
 \end{equation}
where,
\begin{equation}
\label{rstar}
r^* = r + 2 M \ln(r/(2M)-1)
\end{equation}
These functions are directly related to the gravitational wave amplitudes.
The appropriate expressions are given below. 

The initial data are,
\begin{eqnarray}
\label{initZerLLm0}
 \psi _{2, \,0}|_{t=0} & = &  - {8 \sqrt{5 \pi}  r(7 \sqrt{r}+5
\sqrt{r-2M})M \over 15 (\sqrt{r}+\sqrt{r-2M})^5 (2r+3M)} \nonumber \\
 \left. {\partial \psi_{2, \,0} \over \partial t} \right|_{t=0} & = & 0 
\end{eqnarray}
and,
\begin{eqnarray}
\label{initZerLLmpm2}
 \psi _{2, \,\pm2}|_{t=0} & = &   {4 \sqrt{30 \pi}  r(7 \sqrt{r}+5
\sqrt{r-2M})M \over 15 (\sqrt{r}+\sqrt{r-2M})^5 (2r+3M)} \nonumber \\
 \left. {\partial \psi_{2, \,\pm2} \over \partial t} \right|_{t=0} & = & 0 
\end{eqnarray}
and we notice that, since all the $\psi _{2,i}(t,r)$ satisfy the same equation
this implies 
\begin{equation}
\psi _{2,\pm 2}(t,r)= - \sqrt{{3 \over 2}} \; \psi _{2,0}(t,r)
\end{equation}

To compute the wave forms, and radiated angular momentum and energy, we
need to know the asymptotic behavior of the Regge-Wheeler gauge functions.
This is immediately obtained from the knowledge of the asymptotic behavior of
the corresponding Zerilli functions. Proceeding as indicated in
\cite{physrep}, we obtain, 
\begin{equation}
\psi_{2,m}(t,r) = \psi_{2,m}^{(0)}(t-r^*) +\psi_{2,m}^{(1)}(t-r^*)/r+
\psi_{2,m}^{(12}(t-r^*)/r^2+\psi_{2,m}^{(3)}(t-r^*)/r^3+{\cal{O}}(1/r^4)
\end{equation}
where the functions $\psi_{2,m}^{(i)}(x)$  may be written uniquely in terms of
a single function ${\cal{F}}_{2,m}$ in the form
$$
\psi_{2,m}^{(0)}(x) = {d^2 {\cal{F}}_{2,m} \over dx^2} \;\;\;,\;\;\;
\psi_{2,m}^{(1)}(x) =  3 {d {\cal{F}}_{2,m} \over dx} $$
\begin{equation}
\psi_{2,m}^{(2)}(x) = 3{\cal{F}}_{2,m}- 3 M{d{\cal{F}}_{2,m} \over
dx} \;\;\;,\;\;\; 
\psi_{2,m}^{(3)}(x) = -3 M {\cal{F}}_{2,m} + {21 M^2 \over
4} {d{\cal{F}}_{2,m} \over dx}  
\end{equation}
where $m = 0, \pm 2$, and the functions ${\cal{F}}_{2,m}$ are determined by
the initial data, through the evolution equations.

The relation between the functions $\psi_{2,m}$ and the gravitational wave
amplitudes results from their gauge invariance. In an asymptotically flat
gauge we find that, for large $r$, and $t$, we have 
\begin{equation}
\psi_{2,m}(t,r) \simeq  r G_{2,m}(t,r)
\end{equation}
where $\simeq$ implies equality up to terms of order $r^{-1}$.
This result is used in Sections VI and VII, to compute the radiation of energy
and angular momentum.

\subsection{Order $J^2$}

To order $J^2$ we only have perturbations for $\ell=0$, (that contribute only
to the ADM mass), and with $\ell=2$, $m=0$, of even parity, that play a role
in the wave forms and radiated energy. The relevant part of the evolution of
these terms may be obtained in different forms. One way is that considered in
\cite{singlespin}. Here we describe an alternative treatment. As indicated
previously, we fix the first order (order $J$) gauge, to the ``Boyer -
Lindquist - Kerr'' form, $k1_{1,0} (t,r)=0$, $k0_{1,0} = 4\sqrt{(\pi/3)}
J/r$. This implies  a second order (order $J^2$) gauge transformation on the
original Bowen - York order $J^2$ data that, as indicated, introduces only
a (time independent) change in the lapse function $H0$.

Writing the second order Einstein equations it is easy to prove that the 
Zerilli - Moncrief function defined by,
\begin{eqnarray}
 \label{Zerfunl2JJ}
\chi^{J^2} _{2, \,0}(t, \,r) & = & {r(r-2M) \over 3 (2r+3M)}\left[{H2}_{2,
\,0}- r {\partial {K}_{2, \,0}   \over \partial r} - {r-3M \over r-2M} {K}_{2,
\,0}\right] \nonumber \\
& & + {r^2 \over (2r+3M)} \left[{K}_{2, \,0}+(r-2M) \left( {\partial {G}_{2,
\,0}   \over \partial r} - {2 \over r^2} {h1}_{2, \,0}\right) \right]
\end{eqnarray}
where $H2$, $h1$, $K$, and $G$ are the corresponding Regge-Wheeler
coefficients of order $J^2$, obeys the equation, 
\begin{eqnarray}
\label{Zereql2JJ}
{\partial^2  \chi^{J^2} _{2, \,0} \over \partial t^2} 
& = & {\partial^2 \chi^{J^2} _{2,
\,0} \over \partial {r^*}^2}- 6 {(r - 2 M)(3 M^3 + 6 r M^2 + 4r^2 M +
4r^3)\over r^4 (2 r + 3 M)^2}\,\chi^{J^2} _{2, \,0} \nonumber \\
& & -{16 \sqrt{5 \pi} (r-2M)(9M^2+19rM+11r^2) \over 5 (2r+3M)^2 r^6}
 \end{eqnarray}
where the second line contains the ``source'' terms, corresponding to the
quadratic contributions of the order $J$ perturbations. To establish the
relation between this function and the gravitational wave amplitude we
consider the asymptotic behavior of the perturbations for large $t$, and $r$. 
To this end we expand the perturbations in sums of terms of the form
$f_n(t-r^*)/r^n$, and obtain relations between the functions $f_i$, as in
\cite{physrep}. The results, restricted to leading orders in $1/r$, may be
written in the form, 
\begin{eqnarray}
\label{asympl2JJ}
 {H2}_{2, \,0} & = & 12 {G_0(t-r^*) \over r^3} + O(1/r^4) \nonumber \\
{h1}_{2, \,0}& = &  - 2 {{G'}_0(t-r^*) \over r} + O(1/r^2) \nonumber \\
{G}_{2, \,0} & = & {{G''}_0(t-r^*) \over r} + 2 {{G'}_0(t-r^*) \over r^2}
+O(1/r^3)  \nonumber \\   
 {K}_{2, \,0} & = & 3 {{G''}_0(t-r^*) \over r} + 6 {{G'}_0(t-r^*) \over r^2}
+O(1/r^3) 
\end{eqnarray}
where $G_0(x)$ is a function of a single variable, determined by the initial
data through the evolution equations, and a prime indicates a derivative.
Replacing these expansions in (\ref{Zerfunl2JJ}), we find, 
\begin{equation}
\label{chil2JJas} 
\chi^{J^2} _{2, \,0}(t, \,r) = {G''}_0(t-r^*) + { 3 {G'}_0(t-r^*)
\over r} +O(1/r^2)
\end{equation} 

Therefore, asymptotically, to leading order in $1/r$ we have $ \chi^{J^2} _{2,
\,0}(t, \,r) \simeq  r {G2}_{2, \,0} (t,r)$, and $ \chi^{J^2} _{2, \,0}$ may be
identified directly with the gravitational wave amplitude. The initial data
for solving (\ref{Zereql2JJ}) is obtained from the results of Section II.
After replacement and simplification, we find,
\begin{eqnarray}
\label{initZerl2JJ}
 \chi^{J^2} _{2, \,0}|_{t=0} & = &  - {4 \sqrt{5 \pi} (6 r - 5 M) \over 75 r^2 M
(2r+3M)} \nonumber \\
 \left. {\partial \chi^{J^2} _{2, \,0} \over \partial t} \right|_{t=0} & = &
 0 
\end{eqnarray}

This is used in Section VI to obtain the indicated results.
 
\subsection{Order $J\, L^2$, $\ell=2$, even parity, perturbations}

At order $J\, L^2$ we have $\ell=2$, even, and $\ell=3$, odd parity,
contributions. Formally, these are of second
order, since we are considering both $J$, and $L^2$ as first order, and they
satisfy linear inhomogeneous equations, with source terms bilinear in the
order $J$, and order $L^2$ terms. Now, given any solution of these equations,
together with those of order $J$, and $L^2$, it is always possible to perform
first a first order gauge transformation that takes the order $J$ ($\ell=1$,
odd parity) terms to a standard form, that we have taken as the
Boyer-Lindquist-Kerr form, given in Section III, and the order
$L^2$ ($\ell=2$, even parity) terms, to the standard Regge-Wheeler
gauge, where one may write the Regge-Wheeler gauge functions uniquely
in terms of the Zerilli function, $\psi$, (as in (\ref{ele2L2})), and
the order $L^2$ Einstein equation are satisfied if $\psi$ satisfies
the Zerilli equation.

After the previous gauge transformations, a third gauge transformation may be
used to put the $\ell=2$, even parity, order $J L^2$ terms, also in a standard
Regge - Wheeler gauge form. In this gauge we have that one of the Einstein
equations takes the form,
\begin{equation}
H0_{2,2} = H2_{2,2}- {8 i \over 3 (r-2M)} {\partial K^{L^2}_{2,2} \over
\partial t} \end{equation}
where $K^{L^2}_{2,2}$ is of order $L^2$, and all other Regge-Wheeler
coefficients are of order $JL^2$. Taking this into account, it is
straightforward to prove that, as in the cases considered in \cite{physrep},
one can construct an infinite family of generalizations of the Zerilli
function, that satisfy a Zerilli type inhomogeneous equation. From this family
we have singled out the following function: 
\begin{equation}
\label{zerfunJL2}
\chi_{2,2} = {r (r-2M) \over 3 (2r+3M)}\left[H2_{2,2}-r \partial_r K_{2,2}
\right] + {r \over 3} K_{2,2}   -  { 2 i   \over   (2r+3M)} \partial
_t \psi_{2,2} 
\end{equation}

It satisfies the equation,
\begin{eqnarray}
\label{zereqJL2}
{\partial^2 \chi_{2,2} \over \partial t^2}  & = & {\partial^2 \chi_{2,2}\over
\partial {r^*}^2}  - 6 \left(1-{2M \over
r}\right){(4r^3+4Mr^2+6rM^2+3M^3) \over r^3 (2r+3M)^2}\chi_{2,2} \nonumber \\ 
& & -{8 i (4r^3+56r^2M+36rM^2+15M^3) \over r^3 (2r+3M)^3 }{\partial 
\psi_{2,2} \over \partial t}   
\end{eqnarray}
 
The corresponding inversion formulas are,
\begin{eqnarray}
\label{inversion2}
 H2_{2,2} & = &  {(2r^2-2rM+3M^2) \over r (2r+3M)} {\partial \chi_{2,2}\over
\partial r}   -{3 (3M^3+6M^2r+4Mr^2+4r^3) \over r^2(2r+3M)^2} \chi_{2,2}
\nonumber \\ 
& & +(r-2M) {\partial^2 \chi_{2,2}\over \partial
r^2}+2i{(8r^4-60M r^3-12M^2r^2+15M^3r +27M^4) \over
r^2(r-2M)(2r+3M)^3)} {\partial \psi_{2,2}\over \partial t}  
\nonumber \\  & & +2 i {(2r^2+16rM+9M^2)
\over 3 r (2r+3M)^2)} {\partial^2 \psi_{2,2}\over \partial r \partial t} 
\nonumber \\ 
 H0_{2,2} & = &  H2_{2,2}  -{8 i (r^2+rM+M^2) \over
r^2(r-2M)(2r+3M)}{\partial \psi_{2,2}\over \partial t}  
+ {4 i \over 3 r} {\partial^2 \psi_{2,2}\over \partial r \partial t} 
\nonumber \\ 
H1_{2,2} & = & {(2r^2-6rM-3M^2) \over (r-2M)(2r+3M)} {\partial \chi_{2,2}\over
 \partial t} + r {\partial^2 \chi_{2,2}\over \partial r \partial t} 
\nonumber \\ 
& & +{2 i (16r^5-192Mr^4-336M^2r^3-564M^3 r^2 -486 M^4 r -135 M^5)
\over r^4 (2r+3M)^4} \psi_{2,2}
 \nonumber \\ 
& & +{2 i (16 r^3+28 Mr^2+78 M^2 r +45 M^3) \over 3 r^3 (2r+3M)^2}
{\partial \psi_{2,2}\over \partial r} 
\nonumber \\ 
& & + {4 i (r-2M)(9M^2+18Mr +r^2) \over 3 r^2 (2r+3M)^2} {\partial^2
\psi_{2,2}\over \partial r^2} 
\nonumber \\ 
K_{2,2} & = &  {6(r^2+rM+M^2) \over r^2 (2r+3M)}\chi_{2,2} + \left(1-{2M \over
 r} \right) {\partial \chi_{2,2}\over
\partial r}
+{2 i (2r^2+16rM+9M^2) \over r^2 (2r+3M)^2}{\partial
\psi_{2,2}\over \partial t}  
\end{eqnarray}

Again, to compute the asymptotic behavior of the Regge-Wheeler gauge
functions. we need to find the asymptotic behavior of the corresponding
Zerilli functions. Proceeding again as  in \cite{physrep}, we obtain,  
\begin{equation}
\chi_{2,2}(t,r) = \chi_{2,2}^{(0)}(t-r^*) +\chi_{2,2}^{(1)}(t-r^*)/r+
\chi_{2,2}^{(2)}(t-r^*)/r^2+\chi_{2,2}^{(3)}(t-r^*)/r^3+{\cal{O}}(1/r^4)
\end{equation}
where now we have,
\begin{eqnarray}
\chi_{2,2}^{(0)}(x) & = & {d^2 {\cal{G}}_{2,2} \over dx^2}  \nonumber \\
\chi_{2,2}^{(1)}(x) & = & 3 {d {\cal{G}}_{2,2} \over dx} \nonumber \\
\chi_{2,2}^{(2)}(x) & = & 3{\cal{G}}_{2,2}- 3 M{d{\cal{G}}_{2,2} \over
dx} +  i {d^2 {\cal{F}}_{2,2} \over dx^2}\nonumber \\
\chi_{2,2}^{(3)}(x) & = & -3 M {\cal{G}}_{2,2} + {21 M^2 \over
4} {d{\cal{G}}_{2,2} \over dx} + 2 i {d{\cal{F}}_{2,2} \over dx} +
{23 i M \over 3} {d^2 {\cal{F}}_{2,2} \over dx^2} 
\end{eqnarray} 
and similar expressions for $\chi_{2,-2}$. The functions ${\cal{G}}_{2,2}$
are uniquely determined by the initial data. This is obtained starting with
the Bowen - York initial data, and first performing a gauge transformation of
the order $J$, $\ell=1$, odd parity perturbations to a Boyer-Lindquist-Kerr
gauge on the whole initial data, up to order $J L^2$. This does not modify the
order $L^2$ terms but induces a (linear in $J$) change in the order $J L^2$
terms, and a second order gauge transformation on the order $J^2$ terms. On
this modified data we perform a second gauge transformation, this time on the
order $L^2$, $\ell=2$, even parity perturbations to carry it to the Regge -
Wheeler gauge. This induces a further (linear) transformation on the order $J
L^2$ perturbations. Actually, we do not need this last transformation because
the homogeneous part of (\ref{zerfunJL2}), (the part independent of
$\psi_{2,2}$) is gauge invariant under gauge transformations of order $J L^2$
of the $\ell=2$, even parity perturbations, and, therefore, we may obtain that
part of the initial data without explicitly going to the Regge - Wheeler gauge.
After some simplifications, the initial data for $\chi_{2,2}(t,r)$ may be
written in form, 
\begin{eqnarray} \chi_{2,2}|_{t=0} & = & 0 \nonumber \\
\left. {\partial \chi_{2,2} \over  \partial t} \right|_{t=0} & = &
- {8 i \sqrt{30\pi} (16 r^2+57 r M +18M^2 -15 M \sqrt{r}\sqrt{r-2M})
\over 15 (\sqrt{r}+\sqrt{r-2M})^5 (2 r +3 M)^2 r^{5/2}}.
 \end{eqnarray}

We remark that the terms with $\ell=2$, $m=0$, even parity, of order $J\;L^2$
decouple from the order $J$ perturbations, and satisfy homogeneous equations.
Since the initial data for these terms vanish, they vanish and make no
contributions to the evolution of the system.

For the computation of the
radiated energy and wave forms, we need the asymptotic behavior of the
perturbations in an asymptotically flat gauge. It can be checked that in
such a gauge, for large $r$, and $t$, we have
\begin{equation}
\chi_{2,2}(t,r) \simeq  r G_{2,2}(t,r)
\end{equation}
where $\simeq$ implies equality up to terms of order $r^{-1}$, and therefore
$\chi_{2,2}(t,r)$ is directly related to the gravitational wave
amplitude. Similar results hold for $\chi_{2,-2}(t,r)$. These are
used in Sections XX and XX, to compute the radiation of energy and
angular momentum.

\subsection{Order $J L^2$, $\ell=3$, $m= \pm 2$,  odd parity, perturbations}

In this case we may consider again transformations that carry the metric to
the Regge-Wheeler gauge form. The order $J L^2$, $\ell=3$, odd parity,
perturbations satisfy linear equations with a ``source term'' that is linear
in the order $L^2$, $\ell=2$, even parity, perturbations. In the
Regge-Wheeler gauge these may be expressed entirely in terms of the
corresponding Zerilli function. One may in this case verify that if the
Einstein equations are satisfied, then the Regge-Wheeler function defined by,
\begin{equation}
Q_{3,2} (t,r) = -{r \over 10} \left[{2 \over r} k0_{3,2}- {\partial
k0_{3,2} \over \partial r} + {\partial
k1_{3,2} \over \partial t} \right] - {\sqrt{7} \over 105} \left[ {\partial
K^{L^2}_{2,2} \over \partial r} +{3\over r} K^{L^2}_{2,2} \right]
\end{equation} 
where $K^{L^2}_{2,2}$ is of order $L^2$, and all other Regge-Wheeler
coefficients are of order $JL^2$, satisfies the equation,
\begin{equation}
\label{RW32}
{\partial^2 Q_{3,2} \over \partial t^2} - {\partial^2 Q_{3,2} \over \partial
{r^*}^2}+  {6 (r-2M)(2r-M) \over r^4} Q_{3,2}+ {4 (r-2M)\over \sqrt{7} r^4 }
K^{L^2}_{2,2} = 0
\end{equation} 
where
\begin{equation}
K^{L^2}_{2,2} =  {6(r^2+rM+M^2) \over r^2 (2r+3M)}\psi_{2,2} + \left(1-{2M \over
 r} \right) {\partial \psi_{2,2}\over
\partial r} 
\end{equation} 

Using these, and the Einstein equations, one finds the following ``inversion''
formulas,
\begin{eqnarray}
k0_{3,2} & = &  (r-2M) {\partial Q_{3,2} \over \partial r}+ {(r-2M) \over r}
Q_{3,2} + {2 \over 3 \sqrt{7} r} K^{L^2}_{2,2} \nonumber \\
k1_{3,2} & = & {r^2 \over (r-2M)} {\partial Q_{3,2} \over \partial t}
\end{eqnarray}

The asymptotic behavior of $Q_{3,2}$ for large $r$, and $t$, is easily
obtained from (\ref{RW32}). It can written in the form,
\begin{equation} 
Q_{3,2} \simeq {\cal{Q}}_0(t-r^*) + {\cal{Q}}_1 (t-r^*) {1 \over r} 
{\cal{Q}}_2 (t-r^*) {1 \over r^2}  + {\cal{O}}(1/r^3)
\end{equation}
where 
\begin{eqnarray}
{\cal{Q}}_0(t) & = &  {d^2 {\cal{Q}}(t) \over d t^2} \nonumber \\
{\cal{Q}}_1(t) & = &  6 {d {\cal{Q}}(t) \over d t} \nonumber \\
{\cal{Q}}_2(t) & = & 15 {\cal{Q}}(t) - {3  M\over 2 }{d {\cal{Q}}(t) \over d t}
- {\sqrt{7} \over 21} {\partial^2 {\cal{K}}(t) \over \partial t^2} 
\end{eqnarray}
with ${\cal{Q}}$ a certain function that depends on the initial data, and
${\cal{K}}$ related to $\psi_{2,2}$ by, $\partial^2 {\cal{K}} (t-r^*) /
\partial t^2 \simeq \psi_{2,2}(t,r)$, for large $r$, and $t$.

Using these results, and the gauge transformation that relates the
Regge-Wheeler gauge to an asymptotically flat gauge we find that for large
$r$, and $t$, we have,
\begin{equation}
k2_{3,2}(t,r) \simeq - 2 r Q_{3,2}(t,r) \simeq -2 r {\cal{Q}}_0(t-r^*)
\end{equation}
 
The solution for $m =-2$ is the complex conjugate of that for
$m=2$. Therefore, solving (\ref{RW32} we find the gravitational
wave amplitude for this mode.

The foregoing construction gives the complete solution of the perturbative
Einstein equations for the mode considered. For computational purposes,
however, if one is interested only in finding the radiated energy and wave
forms, it is simpler to notice that, after fixing the gauge for the order $J$
(``Kerr'' gauge), and for the order $L^2$ (Regge-Whheler gauge) perturbations,
in a general gauge for the order $J L^2$ perturbations, the function,
\begin{equation}
\widetilde{Q}_{3,2} (t,r) = {(r-2M)\over r^2} \left[ \widetilde{k1}_{3,2}+
{r^2 \over 2}   {\partial \over \partial r} \left({\widetilde{k2}_{3,2} \over
r^2} \right) \right] 
\end{equation} 
satisfies the inhomogeneous Regge-Wheeler equation,
 \begin{eqnarray}
\label{RW30b}
{\partial^2 \widetilde{Q}_{3,2} \over \partial t^2} & = & {\partial^2
\widetilde{Q}_{3,2} \over \partial {r^*}^2} - {6 (r-2M)(2r-M) \over r^4}
\widetilde{Q}_{3,2} \nonumber \\
&  &  -{4 (r-2M)^2 \over \sqrt{7} r^5} \left[ {\partial^2 \psi_{2,2} \over
\partial t \partial r} + {6 (r^2+rM+M^2) \over r(r-2M)(2r+3M)}{\partial
\psi_{2,2} \over \partial t } \right]
\end{eqnarray}

Asymptotically for large $r$, and $t$, $\widetilde{Q}_{3,2}$ admits the
expansion,
\begin{equation}
\widetilde{Q}_{3,2}(t,r) = \widetilde{{\cal{Q}}}_0(t-r^*) + {\cal{O}}(r^{-1})
\end{equation}
 
We remark that, after fixing the gauges for the lower order
perturbations, $ \widetilde{Q}_{3,2}$ is gauge invariant under transformations
of order $J L^2$. In particular, in an asymptotically flat gauge for these
perturbations we have,
\begin{equation}
\widetilde{k2}_{3,2}(t,r),_r \simeq  2 r \widetilde{Q}_{3,2}(t,r) \simeq -2 r
\widetilde{{\cal{Q}}}_0(t-r^*) 
\end{equation}
and, therefore, 
\begin{equation}
\widetilde{k2}_{3,2}(t,r),_t \simeq  - 2 r \widetilde{Q}_{3,2}(t,r)  
\end{equation}
where $\simeq$ implies equality up to terms that decrease as $1/r$ for
large $r$. This last equation establishes the relation between
$\widetilde{Q}_{3,2}$ and the gravitational wave amplitude.

To obtain the initial data for $\widetilde{Q}_{3,2}$, we start with the
Bowen-York initial data, and perform firt a first order transformation on the
order $J$, $\ell=1$, odd parity perturbations, that take these to the
''Kerr'' form, and then, on the result, we perform an order $L^2$
transformation on the $\ell=2$, even parity, to carry these to the
Regge-Wheeler gauge.  Using the resulting form for the order $J L^2$,
$\ell=3$, odd parity perturbations, we find that the initial data for
$\widetilde{Q}_{3,2}$ may be written in the form 
\begin{eqnarray}
\widetilde{Q}_{3,2}(t,r)|_{t=0} & = &  0 \nonumber \\
\left. {\partial \widetilde{Q}_{3,2} \over \partial t}\right|_{t=0} & = &   
{4 \sqrt{210 \pi}\,M \,\sqrt{r - 2\,M}\left( 5 \sqrt{r} +
11 \sqrt{r - 2\,M}\right) 
\over  105 \; (\sqrt{r} + \sqrt{r
- 2\,M})^{5}\,r^{(9/2)}}   
\end{eqnarray}
and we have the same initial data for $m=-2$.

\subsection{Order $J L^2$, $\ell=3$, $m= 0$,  odd parity, perturbations}

In this case the Regge-Wheeler function is given by,
\begin{equation}
Q_{3,0} (t,r) = -{r \over 10} \left[{2 \over r} k0_{3,0}- {\partial
k0_{3,0} \over \partial r} + {\partial
k1_{3,0} \over \partial t} \right] - {\sqrt{35} \over 175} \left[  {\partial
K_{2,0} \over \partial r} +{3  \over   r} K^{L^2}_{2,0} \right]
\end{equation} 
where again $K^{L^2}_{2,2}$ is of order $L^2$, and all other Regge-Wheeler
coefficients are of order $JL^2$, and the Regge-Wheeler equation takes the
form, \begin{equation}
\label{RW30}
{\partial^2 Q_{3,0} \over \partial t^2} - {\partial^2 Q_{3,0} \over \partial
{r^*}^2}+  {6 (r-2M)(2r-M) \over r^4} Q_{3,0}+ {12 (r-2M)\over \sqrt{35} r^4 }
K^{L^2}_{2,0} = 0
\end{equation} 
where
\begin{equation}
K^{L^2}_{2,0} =  {6(r^2+rM+M^2) \over r^2 (2r+3M)}\psi_{2,0} + \left(1-{2M \over
 r} \right) {\partial \psi_{2,0}\over
\partial r} 
\end{equation} 
and the ``inversion'' formulas are,
\begin{eqnarray}
k0_{3,0} & = &  (r-2M) {\partial Q_{3,0} \over \partial r}+ {(r-2M) \over r}
Q_{3,0} + {2 \over \sqrt{35} r} K^{L^2}_{2,0} \nonumber \\
k1_{3,0} & = & {r^2 \over (r-2M)} {\partial Q_{3,0} \over \partial t}
\end{eqnarray}

As in the case $m=\pm2$, the asymptotic behavior of $Q_{3,0}$ for large $r$,
and $t$, is easily obtained from (\ref{RW30}). It can written in the form,
\begin{equation} 
Q_{3,0} \simeq {\cal{Q}}_0(t-r^*) + {\cal{Q}}_1 (t-r^*) {1 \over r} 
{\cal{Q}}_2 (t-r^*) {1 \over r^2}  + {\cal{O}}(1/r^3)
\end{equation}
where 
\begin{eqnarray}
{\cal{Q}}_0(t) & = &  {d^2 {\cal{Q}}(t) \over d t^2} \nonumber \\
{\cal{Q}}_1(t) & = &  6 {d {\cal{Q}}(t) \over d t} \nonumber \\
{\cal{Q}}_2(t) & = & 15 {\cal{Q}}(t) - {3  M\over 2 }{d {\cal{Q}}(t) \over d t}
- {1 \over \sqrt{35}} {\partial^2 {\cal{K}}(t) \over \partial t^2} 
\end{eqnarray}
with ${\cal{Q}}$ again a certain function that depends on the initial data, and
${\cal{K}}$ related to $\psi_{2,0}$ by, $\partial^2 {\cal{K}} (t-r^*) /
\partial t^2 \simeq \psi_{2,0}(t,r)$, for large $r$, and $t$.

Just as in the case $m=\pm2$,we find that for large
$r$, and $t$, the solution of (\ref{RW30}) is related to the amplitude in an
asymptotically flat gauge by, 
\begin{equation}
k2_{3,0}(t,r) \simeq - 2 r Q_{3,0}(t,r) \simeq -2 r {\cal{Q}}_0(t-r^*)
\end{equation}
 
Thus, again the solution of (\ref{RW30}) provides the
gravitational wave amplitude for this mode.

As in the previous subsection, if one is interested only in finding the
radiated energy and wave forms, we notice that, after fixing the
gauge for the order $J$ (``Kerr'' gauge), and for the order $L^2$
(Regge-Wheeler gauge) perturbations, in a general gauge for the order $J L^2$
perturbations, the function, 
\begin{equation}
\widetilde{Q}_{3,0} (t,r) = {(r-2M)\over r^2} \left[ \widetilde{k1}_{3,0}+
{r^2 \over 2}   {\partial \over \partial r} \left({\widetilde{k2}_{3,0} \over
r^2} \right) \right] 
\end{equation} 
satisfies the equation,
 \begin{eqnarray}
\label{RW30a}
{\partial^2 \widetilde{Q}_{3,0} \over \partial t^2} & = & {\partial^2
\widetilde{Q}_{3,0} \over \partial {r^*}^2} - {6 (r-2M)(2r-M) \over r^4}
\widetilde{Q}_{3,0} \nonumber \\
&  &  -{12 (r-2M)^2 \over \sqrt{35} r^5} \left[ {\partial^2 \psi_{2,0} \over
\partial t \partial r} + {6 (r^2+rM+M^2) \over r(r-2M)(2r+3M)}{\partial
\psi_{2,0} \over \partial t } \right]
\end{eqnarray}

Asymptotically for large $r$, and $t$, $\widetilde{Q}_{3,0}$ admits the
expansion,
\begin{equation}
\widetilde{Q}_{3,0}(t,r) = \widetilde{{\cal{Q}}}_0(t-r^*) + {\cal{O}}(r^{-1})
\end{equation}
 
We also have in this case that, after fixing the gauges for the lower order
perturbations, $ \widetilde{Q}_{3,0}$ is gauge invariant under transformations
of order $J L^2$. In an asymptotically flat gauge for these
perturbations we have,
\begin{equation}
\widetilde{k2}_{3,0}(t,r),_r \simeq 2 r \widetilde{Q}_{3,0}(t,r) \simeq -2 r
\widetilde{{\cal{Q}}}_0(t-r^*) 
\end{equation}
and, therefore, 
\begin{equation}
\widetilde{k2}_{3,0}(t,r),_t \simeq - 2 r \widetilde{Q}_{3,0}(t,r)  
\end{equation}
which provides the relation between
$\widetilde{Q}_{3,0}$ and the gravitational wave amplitude.

Proceeding as in the case $m=\pm2$, the initial data for $\widetilde{Q}_{3,0}$
may be written in the form 
\begin{eqnarray}
\widetilde{Q}_{3,0}(t,r)|_{t=0} & = &  0 \nonumber \\
\left. {\partial \widetilde{Q}_{3,0} \over \partial t}\right|_{t=0} & = &   
-{8 \sqrt{7 \pi}\,M \,\sqrt{r - 2\,M} \left( 5 \sqrt{r} +
11 \sqrt{r - 2\,M}\right)
\over  35 \; (\sqrt{r} + \sqrt{r
- 2\,M})^{5}\,r^{(9/2)}}   
\end{eqnarray}

\section{Evolution and waveforms}

All the relevant Regge-Wheeler and Zerilli equations were evolved using  
straightforward modifications of the routines used in \cite{GlNiPrPu}. As
indicated there, the evolution is carried out in the Cauchy domain of some
appropriately chosen interval of the $r^*$-axis, and therefore, we do not
need to impose any boundary conditions. To obtain a definite wave form we
consider an extraction point at $r$ of the order of $100 M$, (the exact point
is actually irrelevant, since the amplitudes are independent of $r$, for large
$r$), and register the different amplitudes as functions of integration time.
For the $\ell=2$, orders $J^2$ and $L^2$, even parity terms we obtain the
familiar ``quasinormal ringing'' wave forms. This is also true for the order
$JL^2$, odd parity, $\ell=3$, amplitudes, although the ringdown frequencies
are different from those for $\ell=2$. For the $\ell=2$, even parity
perturbations of order $JL^2$, we obtain a wave form of the same frequency as
that for $\ell=2$ of order $L^2$, but with a different initial shape, and a
noticeable shift in phase. This last feature is essential for the radiation of
angular momentum. As illustrations of these results we include in Figure 1 a
comparison of the wave forms discussed above. Notice that for $\chi$ 
we have indicated the imaginary part (the real part vanishes in this case) 

\vspace{0.5cm}
 %%%%%%%%%%%%%%%%%

\begin{figure}
\centerline{\psfig{figure=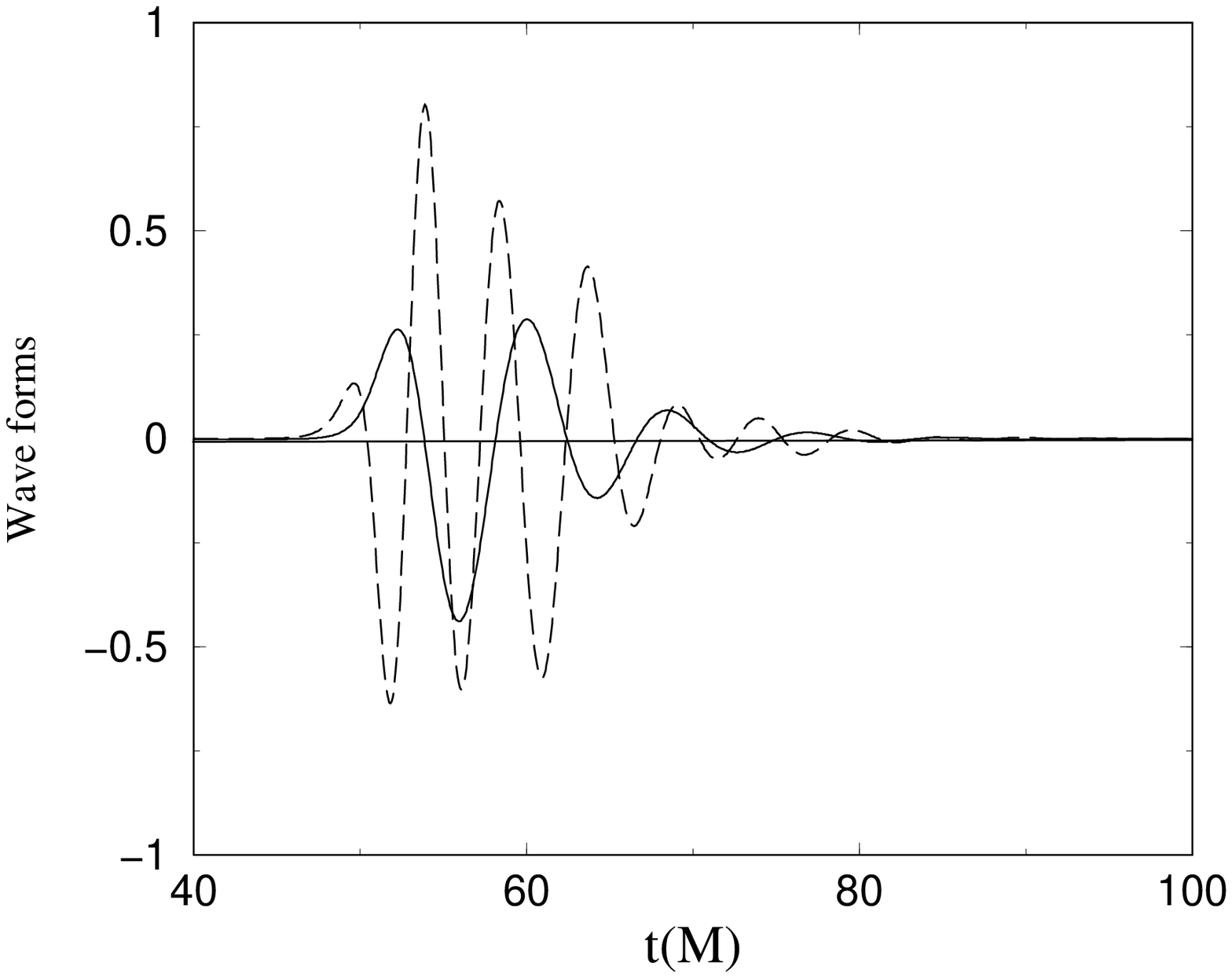,height=50mm}
\hspace{0.5cm}
\psfig{figure=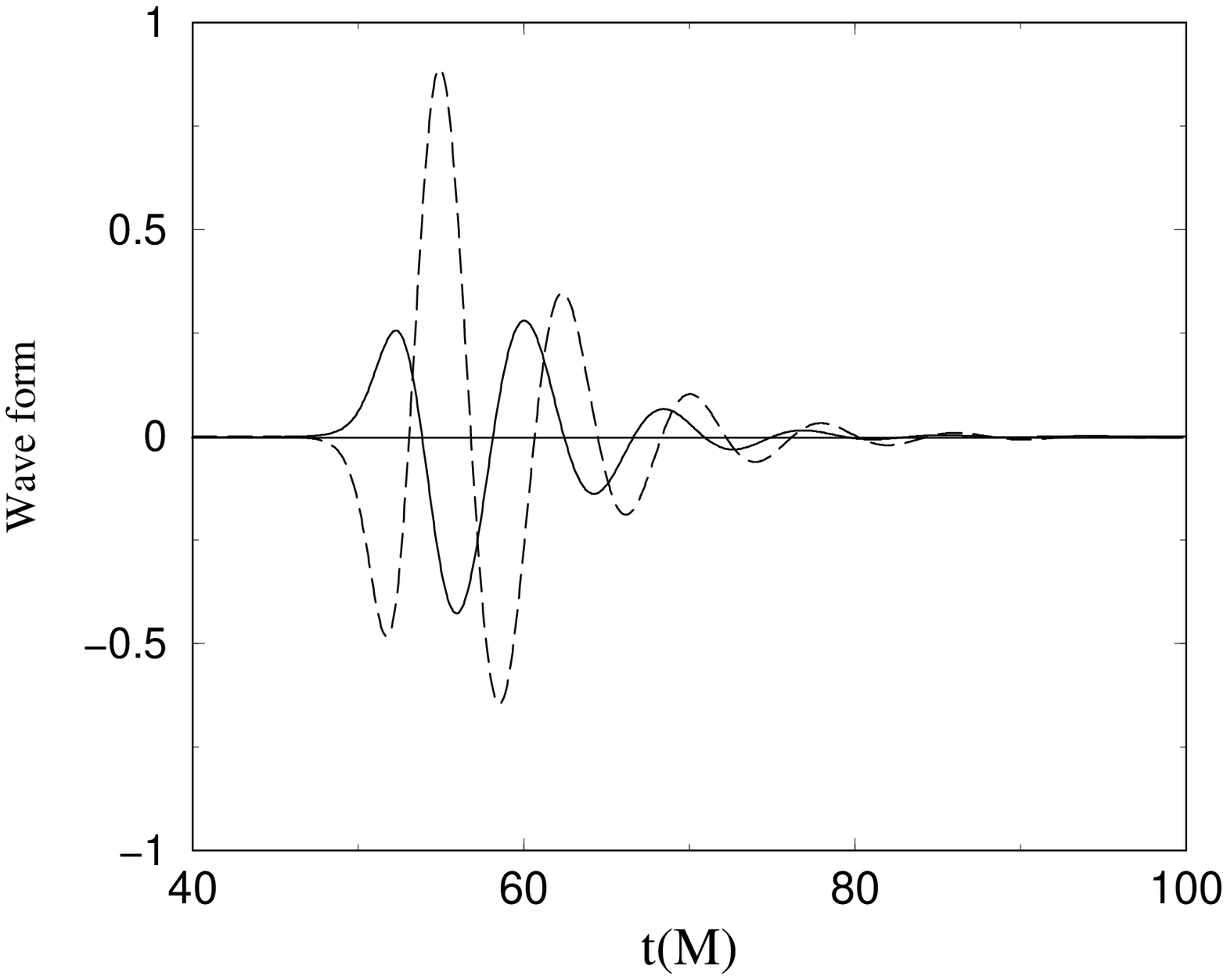,height=50mm}}
\caption{ In the figure on the left the solid curve represents (in arbitrary
units) the wave form corresponding to $\psi_{2,2}$, and the dashed curve
that of $Q_{3,2}$, as functions of time, for fixed $r$. Notice the difference
in ring down frequency, for the different values of $\ell$. The figure on the
right compares, again in arbitrary units, the wave forms corresponding to
$\psi_{2,2}$ (solid curve), and to $\chi_{2,2}$ (dashed curve).
Here we notice the large difference in phase, giving rise to radiation of
angular momentum. }
\end{figure}
 
%%%%%%%%%%%%%%%%%%%%%%%%

\section{Radiated energy}

To compute the power and radiated energy we consider the metric written in a
transverse traceless, asymptotically flat gauge. These conditions are satisfied
asymptotically in what we have called a ``Misner'' gauge in this paper. We may
therefore apply the formula \cite{CPM}

\begin{equation} 
{d \mbox{Power} \over d \Omega} = \lim_{r \rightarrow
\infty}{1 \over 16 \pi r^2} \left[\left({1\over \sin \theta}{\partial
h_{\theta \phi} \over \partial t}\right)^2+ {1 \over 4} \left( {\partial
h_{\theta \theta} \over \partial t}- {1 \over \sin^2 \theta} { \partial
h_{\phi \phi} \over \partial t} \right)^2 \right] 
\end{equation}

Using the general form of the Regge - Wheeler expansion given in
\cite{physrep}, for large $r$, we have,  
\begin{eqnarray}
h_{\theta \theta}-{1 \over \sin^2 \theta} h_{\phi \phi}& = &  r^2
\sum_{\ell=2}^{\infty} \sum_{m=-\ell}^{m=\ell} G_{\ell, m} \left[{\partial^2 
\;   Y_\ell{}^m\over \partial \theta^2} - {1 \over \sin^2 \theta}{\partial^2
Y_\ell{}^m  \over \partial \phi^2} -  \; {\cos \theta \over \sin \theta}\;
{\partial  \; Y_\ell{}^m\over \partial \theta} \right] \nonumber \\ 
& & + 2
\sum_{\ell=2}^{\infty} \sum_{m=-\ell}^{m=\ell} k2_{\ell, m} \left[ {1 \over
\sin(\theta)}{\partial^2 Y_\ell{}^m \over \partial \theta \;
\partial \phi} - {\cos(\theta) \over \sin^2(\theta)} {\partial Y_\ell{}^m
 \over \partial \phi} \right]  \nonumber \\ 
{1 \over \sin \theta} h_{\theta \phi} & = & {1\over 2}
\sum_{\ell=2}^{\infty}\sum_{m=-\ell}^{m=\ell} k2_{\ell, m}   \left[
{\partial^2 \; Y_\ell{}^m \over \sin^2 \theta \partial \phi^2} +
{\cos \theta \over \sin \theta} {\partial  \; Y_\ell{}^m \over
\partial \theta} -  {\partial^2  \; Y_\ell{}^m \over
\partial \theta^2} \right] \nonumber \\  & &  +r^2
\sum_{\ell=2}^{\infty}\sum_{m=-\ell}^{m=\ell} G_{\ell, m} \left[ {\partial^2
Y_\ell{}^m \over \sin \theta \partial \theta \; \partial \phi} -
{\cos \theta  \over \sin^2 \theta} {\partial Y_\ell{}^m \over
\partial \phi} \right]\
\end{eqnarray}
where the sums over $\ell$, and $m$ are restricted to the those of the
perturbation considered. After expansion of the coefficients in the form,
\begin{eqnarray}
G_{\ell, m}(t,r) & = & J^2 G^{(J^2)}_{\ell, m}(t,r)+L^2
G^{(L^2)}_{\ell, m}(t,r)+J L^2 G^{(JL^2)}_{\ell, m}(t,r) \nonumber \\
k2_{\ell, m}(t,r)& = & J L^2 k2^{(JL^2)}_{\ell, m}(t,r) 
\end{eqnarray}
where we have explicitly indicated the orders of the terms, an integration
over the angles gives, 
\begin{eqnarray}
\mbox{Power} & = & \lim_{r \rightarrow \infty} {3 r^2 \over 8 \pi}
\left[ J^4 \left(\dot{G}^{(J^2)}_{2,0}\right)^2
+2 J^2 L^2 \dot{G}^{(J^2)}_{2,0}  \dot{G}^{(L^2)}_{2,0}  
+L^4 \left( \left(\dot{G}^{(L^2)}_{2,0}\right)^2
+2   \dot{G}^{(L^2)}_{2,2} \dot{G}^{(L^2)}_{2,-2}
 \right) \right. \nonumber \\
& & \left. +{5 J^2 L^4 \over r^4} \left(\left(\dot{k2}^{(JL^2)}_{3,0} 
\right)^2 +2   \dot{k}2^{(JL^2)}_{3,2}  
\dot{k}2^{(JL^2)}_{3,-2}  \right) +2 J^2 L^4  
\dot{G}^{(JL^2)}_{2,2}  \dot{G}^{(JL^2)}_{2,-2} \right] 
\end{eqnarray}
where an overdot indicates $\partial /\partial t$. Therefore, in terms of the
corresponding Regge - Wheeler and Zerilli functions, the total radiated energy
is given by,
\begin{eqnarray}
\mbox{Energy} & = & \lim_{r \rightarrow \infty} {3 \over 8
\pi}\int_0^{\infty} \left[ J^4 \left(\dot{\chi}^{(J^2)}_{2,0}\right)^2
+2 J^2 L^2 \dot{\chi}^{(J^2)}_{2,0}  \dot{\psi}^{(L^2)}_{2,0}  
+4 L^4   \left(\dot{\psi}^{(L^2)}_{2,0}\right)^2
 \right. \nonumber \\
& & \left. +{20 J^2 L^4} \left(\left(\dot{Q}^{(JL^2)}_{3,0} 
\right)^2 +2   \dot{Q}^{(JL^2)}_{3,2}  
\dot{Q}^{(JL^2)}_{3,-2}  \right) +2 J^2 L^4  
\dot{\chi}^{(JL^2)}_{2,2}  \dot{\chi}^{(JL^2)}_{2,-2} \right] dt 
\end{eqnarray}

After numerical integration of the evolution equations, with the initial data
derived previously, we obtain
\begin{eqnarray}
\mbox{Energy}/M & = & 7.8 \times 10^{-4} (J/M^2)^4
 -2.8 \times 10^{-5} (J/M^2)^2 (L/M)^2+9.8 \times 10^{-5}(L/M)^4 
\nonumber \\ & &  +1.3 \times 10^{-3} (J/M^2)^2 (L/M)^4(\ell=2) 
+2.1 \times 10^{-5} (J/M^2)^2 (L/M)^4 (\ell=3) 
\end{eqnarray}
where in the second line we have written separately the $\ell =2$, even
parity, and $\ell=3$, odd parity contributions of order $J L^2$.

\section{Radiation of angular momentum}

The lowest order in the radiation of angular momentum comes from the
interference of the $\ell=2$, $m= \pm2$, even parity, outgoing waves of order
$L^2$, with those of order $J L^2$, giving a contribution of order $J L^4$. We
may obtain the expression for the radiated angular momentum using the formulas
in \cite{thorne}. This implies, however, that we need to obtain the
asymptotic transverse traceless gauge amplitudes, which implies a rather
lengthy computation. We may instead use the fact that the following expression,
\begin{equation}
\label{angular1}
{\cal{J}} = \lim_{r \rightarrow \infty} \left[{1 \over 4 \sqrt{3 \pi}}
\left(r^2  k0_{1,0},_r(r,t) - 2 r k0_{1,0}(r,t) r^2
-k1_{1,0},_t(r,t)\right)\right]  
\end{equation}
where $k0_{1,0}(r,t)$, and $k1_{1,0}(r,t)$ represent the corresponding
Regge-Wheeler coefficients in an asymptotic expansion of the metric for large
$r$, corresponds, for $t=0$, and to leading order, to the (ADM) angular
momentum on the initial hypersurface.  It is clear that ${\cal{J}}$, defined by
(\ref{angular1}) is time dependent. Since  ${\cal{J}}(t=0)$ represents the
initial value of the angular momentum of the system, we should consider
$\lim_{t \rightarrow \infty}{\cal{J}}(t)$, as the final value of the angular
momentum, after the system has settled to its final Kerr black hole. The
radiated angular momentum is therefore, 
\begin{equation} 
\label{angular1a}
\Delta {\cal{J}} ={\cal{J}}(0) - \lim_{t \rightarrow \infty}{\cal{J}}(t)
\end{equation}

To compute $\Delta {\cal{J}}$ perturbatively, we expand $k0_{1,0}(r,t)$, 
$k1_{1,0}(r,t)$, in powers of the perturbation parameters $J$, and $L$, 
\begin{eqnarray}
k0_{1,0}(r,t) & = & J {\;} k0^{(1)}_{1,0}(r,t) +J L^2 {\;}k0_{1,0}^{(2)}(r,t)
+J L^4 {\;}k0_{1,0}^{(3)}(r,t) +J^2 L^4 {\;}k0_{1,0}^{(4)}(r,t) +  \dots
\nonumber \\  k1_{1,0}(r,t) & = & J {\;} k1^{(1)}_{1,0}(r,t) +J L^2
{\;}k1^{(2)}_{1,0}(r,t)+J
L^4 {\;}k1_{1,0}^{(3)}(r,t) +J^2 L^4 {\;}k1_{1,0}^{(4)}(r,t)+ \dots  
\end{eqnarray}
Notice that there are no contributions of order $J^2$, $L^2$, and $L^4$. If we
consider now the Einstein equations for the different orders, we have that, to
order $J$,
\begin{equation}
\label{angular2}
 {\cal{J}}^{(1)} =  \left[{1 \over 4 \sqrt{3 \pi}} \left(r^2 
{\;}k0^{(1)},_r(r,t) - 2 r k0^{(1)}(r,t) -r^2
k1^{(1)},_t(r,t)\right)\right]  
\end{equation}
is a gauge invariant constant, that, on account of the form of the initial
data, is equal to ${\cal{J}}$ on the initial hypersurface. 

For the higher order terms, a simple, but rather lengthy calculation
shows that, we have, 
\begin{equation}
\label{angular4}
 {\partial \over \partial t} \left[r^2
k0^{(i)},_r(r,t) - 2 r
 k0^{(i)}(r,t) - r^2
k1^{(i)},_t(r,t)\right] = {\cal{S}}^{(i)} 
\end{equation}
where ${\cal{S}}^{(i)}$ is a `source', that depends on the lower order
perturbations. We then have,
\begin{equation}
\label{angular1b}
\Delta {\cal{J}}
=-\lim_{r \rightarrow \infty}\int_0^{\infty}\left[{\cal{S}}^{(2)}+{{\cal{S}}^{(3)}+
\cal{S}}^{(4)}+\dots\right] dt 
\end{equation}

One can show that the net contribution from ${\cal{S}}^{(2)}$ vanishes. Then,
the leading contribution to $\Delta {\cal{J}}$ comes from ${\cal{S}}^{(3)}$.
From Einstein's equations we find that ${\cal{S}}^{(3)}$ is bilinear in the
$\ell=2$,$m=\pm2$, order $L^2$ and order $J L^2$ contributions. Writing
these perturbations in the Regge - Wheeler gauge, and using the expansions of
the previous Section, a straightforward computation shows that,
\begin{equation}
\label{angular1d}
\Delta {\cal{J}}
= {3 i  \over 4 \pi}\lim_{r \rightarrow \infty}\int_0^{\infty}\left[
{\partial \chi_{2,2} \over \partial t} \psi_{2,-2} 
-{\partial \psi_{2,-2} \over \partial t} \chi_{2,2} 
-{\partial \chi_{2,-2} \over \partial t} \psi_{2,2} 
+{\partial \psi_{2,2} \over \partial t} \chi_{2,-2} 
\right] dt
\end{equation}

Using now the results of the numerical evolution, we finally find,
\begin{equation}
\label{angular1e}
{\Delta {\cal{J}} \over {\cal{J}}} = 0.0023 \left({L \over
M}\right)^4  \end{equation}

This result implies that less than 0.3\% of the total angular
momentum will radiated even in the extreme case $L \simeq M$. It is
in good agreement with the corresponding result found in
\cite{psinsp} for the ``pseudo-inspiral'' case.  

 \section{Final comments and conclusions}

In this paper we have numerically evolved a particular set of initial
data, and obtained quantitative expressions for the radiated energy and angular
momentum, for a system of two equal mass rotating black holes, in the close
limit. If we analyze in more detail the expression for the radiated energy, we
find that a large contribution comes from the order $J^2$ amplitude. This is
probably due to the fact that the initial data for rotating black holes
obtained with the Bowen and York prescription, even for a single black hole
\cite{singlespin}, contains a fair amount of ``spurious'' radiation that must
be radiated away before the system can settle to the final Kerr form. One
might then expect that they would be absent, or make a much less significant
contribution if the evolution started with initial data closer to that of a
Kerr black hole. On the other hand, it seems reasonable to assume that the
contributions of the terms of order $L^2$, and $J L^2$ should be much less
dependent on the form of the initial data, since for order $L^2$ they are
independent of spin, and for order $J L^2$ they require the order $J$ data,
and this is ``exactly'' of the Kerr form. It is precisely these terms 
that provide the leading contribution to the radiated angular momentum. This
contribution is such as to decrease the angular momentum of the system. The
radiation may therefore be considered as providing a ``torque'' somewhat in
the manner envisioned in \cite{PriWhe}.

\section*{Acknowledgments}

We are grateful to Manuel Tiglio for his help in the early stages of
this work and  helpful criticism and comments. We are also grateful to
Jorge Pullin and Gaurav Khanna for their comments. This work was
supported in part by grants of the National University of C\'ordoba, CONICET,
Fundaci\'on Antorchas,  Agencia C\'ordoba Ciencia and grant NSF-PHY-9800973.
R.J.G. is a research member of CONICET  (Argentina). A.E.D. is a
CONICET postdoctoral fellow.

%%%%%%%%%%%%%%%%%%%%%%%

\end{document}